\documentclass{PoS}

\title{Encoding field theories into gravities}

\ShortTitle{Encoding field theories into gravities}

\author{\speaker{Sinya Aoki}\\
        Yukawa Institute for Theoretical Physics, Kyoto University, Kyoto 606-8502, Japan \\
        Center for Computational Sciences, University of Tsukuba, Ibaraki 305-8577, Japan \\
        E-mail: \email{saoki@yukawa.kyoto-u.ac.jp}}

\author{Kengo Kikuchi\\
        Yukawa Institute for Theoretical Physics, Kyoto University, Kyoto 606-8502, Japan \\
        E-mail: \email{kengo@yukawa.kyoto-u.ac.jp}}
\author{Tetsuya Onogi\\
	Department of Physics, Osaka University, Osaka 560-0043, Japan\\
       E-mail: \email{onogi@het.phys.sci.osaka-u.ac.jp}}
	
\abstract{We propose a method to give a $d+1$ geometry from  a $d$ dimensional quantum  field theory  
in  the large $N$ expansion. We first construct  a $d+1$ dimensional  field  from the $d$ dimensional one using the gradient flow equation, whose flow time $t$ represents the energy scale of the system such that $t\rightarrow 0$ corresponds to the ultra-violet (UV) while $t\rightarrow\infty$ to the infra-red (IR).  
We define the induced metric using $d+1$ dimensional field operators.    
We show that the metric defined in this way becomes classical in the large $N$ limit:  quantum fluctuations of the metric are suppressed as $1/N$ due to the large $N$ factorization property.
As a concrete example, we apply our method to the O(N) non-linear $\sigma$ model in two dimensions.
We calculate  the three dimensional induced metric, which describes  an AdS  space in the massless limit. 
We finally discuss several open issues  for future investigations.
}

\FullConference{The 33rd International Symposium on Lattice Field Theory\\
		14 -18 July 2015\\
		Kobe International Conference Center, Kobe, Japan*}

\begin{document}

\section{Introduction}
The AdS/CFT (or more generally Gravity/Gauge theory) correspondence\cite{Maldacena:1997re}, which is one of the most surprising and significant findings in quantum field theories, claims a $d$ dimensional conformal field theory is equivalent to some $d+1$ dimensional (super-)gravity theory on the AdS background.
There appeared many evidences for this correspondence after this proposal.
However, this equivalence is still mysterious, though the open/closed string duality may give a proof of the correspondence.  

In  this report, based on our previous publication\cite{Aoki:2015dla}, we investigate such gravity/field theory correspondences from a little different point of view, and propose a general method to derive a geometry from a field theory.
In our method, we extend a $d$ dimensional $N$ component quantum field to a $d+1$ dimensional one via the gradient flow\cite{Narayanan:2006rf,Luscher:2010iy, Luscher:2009eq,Luscher:2013vga}, where the flow time $t$ is regarded as an additional coordinate and corresponds to the energy scale of the original $d$ dimensional theory. Interpreting this $d+1$ dimensional field as the map from $d+1$ dimensional space to the curved manifold in $N$ dimensions, we  define the induced metric in $d+1$ dimensions, which contains information of the $d$ dimensional field theory and its scale dependence. This method is so general that it can be applied to all quantum field theories in principle.
Moreover, the metric defined in this method becomes classical in the large $N$ limit.

This report is organized as follows. We first give a general procedure to give the induced metric from an arbitrary field theory, and discuss some properties of this proposal. We then apply this procedure to the two dimensional O(N) non-linear $\sigma$ model. We evaluate the vacuum expectation value (VEV) of the induced metric in three dimensions, which describes an AdS space in the massless limit.  We finally discuss remaining problems of this proposal for future investigations.

\section{Proposal}
Let us consider  an arbitrary field theory with an action $S$ for a field $\varphi^{a,\alpha}(x)$, where $x$ is $d$ dimensional Euclidean coordinate, $a=1,2,\cdots, N$ is the large $N$ index, and $\alpha$ represents other indices such as Lorentz index, so that $h_{\alpha\beta} \varphi^{a,\alpha}(x)\varphi^{b,\beta}(x)$ becomes Lorentz invariant with a constant tensor $h_{\alpha\beta}$. 

We first define a $d+1$ dimensional field $\phi^{a,\alpha}(t,x)$ from this $d$ dimensional  $\varphi^{a,\alpha}(x)$,
using the gradient flow equation\cite{Kikuchi:2014rla} given by
\begin{eqnarray}
\frac{d}{d t} \phi^{a,\alpha}(t,x) &=& - g^{ab}(\phi(t,x))\left. \frac{\delta S}{\delta \varphi^{b,\alpha}(x)}\right\vert_{\varphi \rightarrow\phi} , \qquad \phi^{a,\alpha}(0,x)=\varphi^{a,\alpha}(x), 
\label{eq:GFE}
\end{eqnarray}
where $g^{ab}$ is the metric of the space for the large $N$ index.
Since the mass dimension of $t$ is $-2$ and $t\ge 0$, we can define new variable $\tau =2\sqrt{t}$ (Here the factor 2 makes some latter results simpler. ) and write $d+1$ dimensional coordinate as $z=(\tau,x)\, \in \mathbb {R}^+(=[0,\infty])\times \mathbb{R}^d$ and the field as $\phi^{a,\alpha}(z)$.  

We now give our proposal to define a  $d+1$ dimensional metric as
\begin{eqnarray}
\hat g_{\mu\nu} (z) &: =& g_{ab}(\phi(z)) h_{\alpha\beta} \partial_\mu \phi^{a,\alpha}(z) \partial_\nu \phi^{b,\beta}(z) ,
\label{eq:metric}
\end{eqnarray}
where the constant tensor $h_{\alpha\beta}$\footnote{In general, we might introduce a $z$ dependent tensor $h_{\alpha\beta}(z)$, but  we only consider the constant case here.}
has a mass dimension $-2(1+d_\varphi)$ with $d_\varphi$ being that for $\varphi$, in order to make the metric dimensionless. This is an induced metric of a $d+1$ dimensional manifold $\mathbb{R}^+\times \mathbb{R}^d$ into a curved space in $\mathbb{R}^N$ with the metric $g_{ab}$.
Using this definition, we can calculate VEV of $g_{\mu\nu}$'s as
\begin{eqnarray}
\langle \hat g_{\mu\nu}(z) \rangle &:=&  \langle  \hat g_{\mu\nu}(z) \rangle_S , \\
\langle \hat g_{\mu_1\nu_1}(z_1) \hat g_{\mu_2\nu_2}(z_2) \rangle &:=& \langle \hat g_{\mu_1\nu_1}(z_1) \hat g_{\mu_2\nu_2}(z_2) \rangle_S, \\
\langle \hat g_{\mu_1\nu_1}(z_1)\cdots \hat g_{\mu_n\nu_n}(z_n) \rangle &:=& \langle \hat g_{\mu_1\nu_1}(z_1) \cdots \hat g_{\mu_n\nu_n}(z_n) \rangle_S, ~~~~
\end{eqnarray}
where $\langle { O} \rangle_S$ is an expectation value of ${ O}(\varphi)$ in the original $d$ dimensional theory with the action $S$ as
\begin{eqnarray}
\langle { O} \rangle_S &:=& \frac{1}{Z}\int { D}\varphi \, { O}(\varphi)\, e^{-S}, 
\quad Z:= \int { D}\varphi \,  e^{-S} 
\end{eqnarray}
in the large $N$ expansion. 
Note that, even though the ``composite" operator $\hat g_{\mu\nu}(z)$ contains a product of two local operators at the same point $z$,
 $\langle \hat g_{\mu\nu}(z)\rangle $ is finite as long as $\tau\not=0$\cite{Luscher:2011bx}.
This is a reason why the induced metric is defined in $d+1$ dimensions from $\phi$.
The induced metric would severely diverge if it were defined in $d$ dimensions from $\varphi$.

Interestingly, quantum fluctuations of the metric $\hat g_{\mu\nu}$ are suppressed in the large $N$ limit due to the large $N$ factorization.
For example, the two point correlation function of $\hat g_{\mu\nu}$ becomes
\begin{eqnarray}
\langle \hat g_{\mu\nu}(z_1)  \hat g_{\alpha\beta}(z_2) \rangle &=&  
\langle \hat g_{\mu\nu}(z_1)  \rangle \langle\hat g_{\alpha\beta}(z_2) \rangle + O\left(\frac{1}{N}\right),~~
\end{eqnarray}
which shows that the induced metric $\hat g_{\mu\nu}$ is classical in the large $N$ limit, and quantum fluctuations  can be calculated in the large $N$ expansion.
This property is  important for  the geometrical interpretation of 
the metric $\hat g_{\mu\nu}$.  In the large $N$ limit, the VEV of the curvature tensor operator can be obtained directly from the VEV of $\hat g_{\mu\nu}$ as if the theory is classical.

\section{O(N) non-linear $\sigma$ model in two dimensions}
As an explicit example, let us consider the O(N) non-linear $\sigma$ model in two dimensions, 
whose action is given by
\begin{equation}
S=\frac{1}{2g^2}\int d^2x\, \sum_{a,b=1}^{N-1} g_{ab}(\varphi) \sum_{k=1}^2 \left(\partial_k \varphi^a(x) \partial^k\varphi^b(x)\right),
\end{equation}
where 
\begin{equation}
g_{ab}(\varphi) = \delta_{ab} + \frac{\varphi^a\varphi^b}{1-\varphi\cdot\varphi}, \quad
g^{ab}(\varphi) = \delta_{ab} - \varphi^a\varphi^b 
\end{equation}
with $\varphi\cdot\varphi =\sum_{a=1}^{N-1}\varphi^a\varphi^a$.
Here the metric $g_{ab}$ appears in the action, as
the $N$-th component of $\varphi$ is expressed in terms of other fields: $\varphi^N =\pm \sqrt{1-\varphi\cdot\varphi}$.
According to our proposal, the three dimensional metric $g_{\mu\nu}(z)$ can be obtained from this theory.

\subsection{Solution to the gradient flow equation in the large $N$}
The solution to the gradient flow equation, obtained in the previous study\cite{Aoki:2014dxa},
is given in the momentum space as
\begin{eqnarray}
\phi^a(t,p) &=& f(t) e^{-p^2 t} \sum_{n=0}^\infty : X_{2n+1}(\varphi, p,t) :
\end{eqnarray}
 where $X_{2n+1}$ is proportional to $\varphi^{2n+1}$ and is $O(1/N^{2n+1})$.
 The leading order term $X_1$ thus becomes $X_1^a(\varphi,p,t) = \varphi^a(p)$ with
 \begin{eqnarray}
f(t) &=& \frac{1}{\sqrt{1-2\lambda J(t)}}, \quad
J(t) = \int_0^t ds I(s), 
 \quad I(t)=
 \int \frac{d^2q}{(2\pi)^2} \frac{q^2 e^{-2q^2 t}}{q^2+m^2}, 
\end{eqnarray}
where $\lambda = g^2 N$ is the 't Hooft coupling constant, and $m$ is the dynamically generated mass, which satisfies
\begin{equation}
1= \lambda \int \frac{d^2q}{(2\pi)^2} \frac{1}{q^2+m^2} =\frac{\lambda}{4\pi}\log\frac{\Lambda^2+m^2}{m^2}
\label{eq:gap}
\end{equation}
with the momentum cut-off $\Lambda$.  We then obtain
\begin{eqnarray}
f(t) &=&e^{-m^2 t} \sqrt{\frac{4\pi/\lambda}{{\rm Ei}\left(-2t(\Lambda^2+m^2)\right)- {\rm Ei}\left(-2t m^2\right)}},~~~
\end{eqnarray}
where  ${\rm Ei}(x)$ is the exponential integral function defined as
${\rm Ei}(-x) = \int d\,x\, e^{-x}/{x}$.
The 2-pt function, which dominate in the large $N$ limit, can be evaluated as
\begin{eqnarray}
\langle \phi^a(t,x)\phi^b(s,y)\rangle_S &=&  \int  \frac{d^2 q}{(2\pi)^2}  \frac{e^{-q^2 (t+s)} e^{iq(x-y)}}{q^2+m^2} 
\delta_{ab} \frac{\lambda}{N} f(t)f(s)
+O(N^{-2}),
\end{eqnarray}
which is finite even in the $\Lambda\rightarrow\infty$ limit, which is equivalent to  $\lambda\rightarrow 0$ 
from eq.~(\ref{eq:gap}).

\subsection{Induced metric}
Our definition of the induced metric for this model becomes
\begin{eqnarray}
\hat g_{\mu\nu}(z) := h\, g_{ab}(\phi (z) ) \partial_\mu \phi^a(z) \partial_\nu \phi^b (z) ,\qquad
z = (2\sqrt{t},x),
\label{eq:metric_3d}
\end{eqnarray}
where the constant $h$ is introduced so as to make the mass dimension of the metric operator $\hat g_{\mu\nu}(z)$  vanish.
This is the induced metric of a three dimensional manifold $\mathbb{R}^+\times \mathbb{R}^2$ into the $N-1$ dimensional sphere defined by $\phi^a$. 

The VEV of the metric, $g_{\mu\nu}$ does not depend on $x$, due to the translational invariance of the two dimensional O(N) non-linear $\sigma$ model, and it can easily be calculated in the large $N$ limit.
We have $g_{i\tau} (\tau) = g_{\tau i} (\tau) = 0$ for $i=1,2$, while
\begin{eqnarray}
g_{ij} (z) &:=& \langle \hat g_{ij} (z ) \rangle = h \langle g_{ab}(\phi) \partial_i\phi^a(t,x) \partial_j\phi^b(t,x) \rangle 
\simeq
\frac{ {h}}{2} \delta_{ij} \lambda f^2(t) I(t ) = \frac{ h}{2} \delta_{ij} \frac{\dot{f}(t)}{f(t)}
\end{eqnarray}
for $i,j=1,2$, 
where  an identity that $\dot f(t):=\displaystyle d f(t)/d t = \lambda f(t)^3 I(t)$ is used.  
In addition, we have
\begin{eqnarray}
g_{\tau\tau} (\tau) &=&  \frac{\tau^2 h}{4} \left\langle 
\dot \phi^a(t,x) g_{ab}(\phi(t,x)) \dot \phi^b(t,x) 
\right\rangle ,
\end{eqnarray}
which can be evaluated by using the gradient flow equation as
\begin{eqnarray}
g_{\tau\tau} (\tau) 
 &\simeq& \frac{\tau^2  h}{4} \left[ \langle \nabla^2\phi\cdot  \nabla^2\phi\rangle - \langle \phi\cdot  \nabla^2\phi\rangle^2\right]  
 =-\frac{\tau  h}{4}\frac{d}{d \tau} \left(\frac{\dot f}{f}\right) .
\end{eqnarray}
Thus the VEV of  the induced metric turns out to be diagonal as
\begin{equation}
g_{\mu\nu} = \left(\begin{array}{ccc}
B(\tau) & 0 & 0 \\
0 & A(\tau) & 0 \\
0 & 0 & A(\tau) \\
\end{array}
\right),
\qquad 
A(\tau) =  \displaystyle \frac{ h}{2}\frac{\dot f(t)}{f(t)}\Bigr\vert_{t=\tau^2/4},\quad
B(\tau) =   -\frac{\tau}{2} A_{,\tau},
\end{equation}
where $f_{,\tau}$ means the derivative of $f$ with respect to $\tau$.
This $A(\tau)$, hence also $ B(\tau)$, is finite in the limit  $\Lambda\rightarrow\infty$, and $A(\tau)$ is given as
\begin{eqnarray}
A(\tau) &=&- \frac{m^2  {h}}{2}\left[1 +\frac{e^{-\tau^2 m^2/2}}{{\rm E_i}(-\tau^2m^2/2) m^2\tau^2/2 }\right].
~~
\label{eq:A}
\end{eqnarray}

From the VEV of  the metric, we can evaluate VEV's of composite operators such as the Einstein tensor $G_{\mu\nu}(\hat g)$ as $\langle G_{\mu\nu}(\hat g)\rangle = G_{\mu\nu}(\langle\hat g\rangle)$,  due to the large $N$ factorization.
After a little calculation, we have
\begin{equation}
G_{\tau\tau} =\displaystyle \frac{A_{,\tau}^2}{4 A^2},\quad 
G_{ij} =\displaystyle \delta_{ij} \left[ \frac{A_{,\tau\tau}}{2B} - \frac{A_{,\tau} B_{,\tau}}{4B^2} - \frac{A_{,\tau}^2}{4AB}\right],
\qquad G_{i\tau} = G_{\tau i} = 0 .
\end{equation}

\subsection{Massless limit and AdS space} 
In the massless limit ($m\rightarrow 0$),
$A$ and its derivatives  are given by
\begin{eqnarray}
A &\simeq & -\frac{1}{\tau^2} \frac{h}{\log (m^2)} \left[1+O\left( \frac{1}{\log(m^2)}\right)\right], 
\qquad
A_{,\tau} \simeq -2A/\tau,
\qquad
A_{,\tau\tau} \simeq 6 A/\tau^2 .
\end{eqnarray}
We here use the expansion ${\rm E_i}(-x) =\log x +\gamma+ \sum_{n=1}^\infty (-x)^n/(n\cdot n!)$.
In order to have positive and finite $g_{\mu\nu}$ in the massless limit, we take $h=-R_0^2 \log (m^2 R_0^2)$, where  a 
mass dimension of the constant $R_0$ is $-1$.
We thus obtain
\begin{equation}
g_{\tau\tau}  = \displaystyle\frac{R_0^2}{\tau^2},  \qquad
g_{ij}  = \displaystyle\delta_{ij} \frac{R_0^2}{\tau^2},\qquad
\Rightarrow \qquad ds^2 = \frac{R_0^2}{\tau^2}\left[  d\tau^2 + (d\vec{x})^2 \right],
\end{equation}
which describes the Euclidean AdS space. The Einstein tensor indeed becomes
\begin{equation}
G_{\mu\nu} = -\Lambda_0 g_{\mu\nu}, \qquad
\Lambda_0 = - \frac{1}{R_0^2} , 
\end{equation} 
which shows that  the cosmological constant $\Lambda_0$ is negative.
It is interesting and suggestive that the AdS geometry appeared from the conformal field theory defined in the massless limit, which corresponds to the UV fixed point of the original theory.

\subsection{Metric in UV and IR limits} 
 In the short distance (UV) limit  ($m\tau \rightarrow 0$), we have
 $A \simeq  -h/[\tau^2 \log (m^2\tau^2)]$, 
so that
\begin{equation}
g_{\tau\tau} \simeq - h/[\tau^2 \log (m^2\tau^2)], \qquad
g_{ij} \simeq - \delta_{ij}h/[\tau^2  \log (m^2\tau^2)] .
\end{equation}

As briefly discussed for generic cases,
if  we had defined the $d$ dimensional metric directly from the $d$ dimensional field theory as
$\hat g_{ij}^d(x) := g_{ab}( \varphi(x)) \partial_i \varphi^a(x)  \partial_j \varphi^b(x)$,
the VEV of $\hat g_{ij}^d(x)$ would give the UV divergence due to the short distance singularity of $\varphi^a (x) \varphi^b (y)$ at $x\rightarrow y$.
In contrast,  the $d+1$ dimensional metric $\hat g_{\mu\nu}$ defined from the $d+1$ dimensional field via eq.~(\ref{eq:metric}) is UV finite, since the flowed field $\phi (t,x)$ and any local composite operators are expected to be finite as long as $t\sim\tau^2$ is non-zero\cite{Luscher:2011bx,Aoki:2014dxa,Makino:2014sta,Makino:2014cxa}.
This is a reason why the induced metric is defined on $d+1$ dimensions, not on $d$ dimensions.  Consequently, the classical metric, $g_{\mu\nu}$, is UV finite in our proposal. 
Note that the UV divergence in the original two dimensional O(N) non-linear $\sigma$ model appears in the $m\tau \rightarrow 0$ limit as $g_{\mu\nu} \sim 1/(\tau^2\log m^2\tau^2)$.
In the UV limit, the ``effective" cosmological constant becomes  
\begin{equation}
\Lambda_0^{\rm eff} = -\frac{\log (m^2\tau^2)}{R_0^2\log(m^2R_0^2)},
\end{equation} 
where  non-conformal natures of the original two dimensional asymptotic-free field theory appear as its $\log \tau^2$ dependence.

In the $m\tau\rightarrow\infty$ (IR) limit, on the other hand,
we have $A \simeq  h/\tau^2 $, 
which gives
\begin{equation}
g_{\tau\tau} \simeq h/\tau^2, \qquad
g_{ij} \simeq  \delta_{ij}h/\tau^2,
\qquad 
\Lambda_{0}^{\rm eff} = [R_0^2\log(m^2R_0^2)]^{-1}.
\end{equation}
Thus the theory becomes asymptotically  AdS
 ($\Lambda_0^{\rm eff} < 0$) if  $\log(m^2 R_0^2) < 0$.
This result  looks rather non-trivial,
since  we naively expect that $\Lambda^{\rm eff}_0\rightarrow 0$ in the IR limit
due to decouplings of  all massive modes in this limit.

If we assume  the Einstein equation,  $G_{\mu\nu} =8\pi G T_{\mu\nu}$, we can define the energy momentum tensor $T_{\mu\nu}$, which becomes $T_{\mu\nu}=\delta_{\mu\nu}/(8\pi G \tau^2)$ in both UV and IR limits.
Since $T_{\mu\nu}$ does not depend on $m$, this result  holds even at $m=0$.
Therefore, if we define the energy momentum tensor for the matter as $T_{\mu\nu}^{\rm matter}:=T_{\mu\nu} + g_{\mu\nu}\Lambda_0/(8\pi G)$, 
$T_{\mu\nu}^{\rm matter}$ vanishes in both UV and IR limits.

\section{Discussions}
In our method proposed in this report,  a relation between a  geometry and a given field theory is explicit by construction.
Furthermore, the method can be applied to an arbitrary quantum field theory, though the large $N$ expansion is required for geometrical interpretations.
If the theory is solvable in the large $N$ limit, the VEV of $\hat g_{\mu\nu}$ can be  exactly evaluated.
We now discuss  the remaining issues of our proposal. 

First, we need a complete dictionary  to interpret a quantum field theory in terms of the corresponding  metric operator and vice versa.  
The dynamically generated mass  $m$ for the two dimensional O(N) non-linear $\sigma$ model can be extracted 
from the asymptotic behavior of the metric that $g_{\mu\nu}\sim 1/[\tau^2\log(m^2\tau^2)]$ as $m\tau\rightarrow 0$.
This information, however,  gives only a partial knowledge of the field theory.  
We also need a method to determine a structure of  the matter content using $T_{\mu\nu}^{\rm matter}$ we obtained from 
the Einstein equation for the metric.

Secondly,  we should investigate what kind of geometry emerges from
various large $N$ models other than the two dimensional O(N) non-linear $\sigma$ model,
since our method can be applied to a large class of quantum field theories including conformal theories.
One possible direction is to introduce a source term into the original theory to break the translational invariance,
so that the metric can have  nontrivial $z$ dependences. 
In particular, it  is important to find the field theory set-up which induces the black hole geometry. 

Thirdly,  it is interesting to study the fluctuations around the background geometry.
In principle, we can evaluate an arbitrary correlation function for the metric $\hat g_{\mu\nu}$ including
quantum fluctuations  in the large $N$ expansion.
In practice, however, calculations in the next leading order  become much more involved than those in the large $N$ limit\cite{Aoki:2014dxa}.
Although the action for $\hat g_{\mu\nu}$ is absent in our approach, we may effectively define the quantum theory of the metric.
It is interesting to investigate whether this quantum theory  is renormalizable (or even UV finite) or not, in contrast to the unrenormalizable quantum theory of the Einstein gravity.
Complementally,  it is also important  to calculate an effective action for the composite operator
$\hat g_{\mu\nu}$.


Finally, simple choices for the induced metric defined from  gauge theories are given by
\begin{eqnarray}
\hat g_{\mu\nu}(z) &:=& h \sum_{i,j=1}^d{\rm Tr}\ D_\mu F_{ij}(z) D_\nu F^{ij}(z), \\
\hat g_{\mu\nu}(z) &:=& h \sum_{\alpha=0}^d {\rm Tr}\ F_{\mu\alpha}(z)F_{\nu}{}^{\alpha}(z),
\end{eqnarray}
where $D_i$ ($i=1,\cdots,d$) is the covariant derivative in $d$ dimensions, while $D_\tau :=\partial_\tau$,  and then the field strength is given as $F_{\mu\nu}:=[D_\mu,D_\nu]$. 
Both definitions are invariant under the $\tau$-independent gauge transformation\cite{Luscher:2010iy,Luscher:2009eq,Luscher:2013vga}. 
It would be interesting to calculate the induced metric form 
the large $N$ gauge theory in two dimensions ('t Hooft model)\cite{'tHooft:1974hx} using our method.

\bigskip

S.A. would like to thank Drs.  N. Ishibashi, S. Sugimoto, T. Takayanagi and Y. Yokokura for their useful comments. 
 We also thank Dr. S. Yamaguchi for his comment on the AdS metric.
This work was supported by Grant-in-Aid for JSPS Fellows Grant Number 25$\cdot$1336 and by the Grant-in-Aid of the Japanese Ministry of Education (Nos. 25287046, 26400248), by MEXT SPIRE and JICFuS.


\begin{thebibliography}{99}
\bibitem{Maldacena:1997re}
  J.~M.~Maldacena,
  Int.\ J.\ Theor.\ Phys.\  {\bf 38}, 1113  (1999) 
   [Adv.\ Theor.\ Math.\ Phys.\  {\bf 2}, 231 (1998) ]
  [hep-th/9711200].

\bibitem{Aoki:2015dla}
  S.~Aoki, K.~Kikuchi and T.~Onogi,
  PTEP {\bf 2015} (2015) 10,  101B01
  [arXiv:1505.00131 [hep-th]].
  
\bibitem{Narayanan:2006rf} 
  R.~Narayanan and H.~Neuberger,
  JHEP {\bf 0603}, 064 (2006)
  [hep-th/0601210].
      
\bibitem{Luscher:2010iy} 
  M.~L\"uscher,
  JHEP {\bf 1008}, 071 (2010) [JHEP {\bf 1403}, 092 (2014)]
  [arXiv:1006.4518 [hep-lat]].

\bibitem{Luscher:2009eq} 
  M.~L\"uscher,
  Commun.\ Math.\ Phys.\  {\bf 293}, 899 (2010)
  [arXiv:0907.5491 [hep-lat]].

\bibitem{Luscher:2013vga} 
  M.~L\"uscher,
  PoS LATTICE {\bf 2013}, 016 (2014)
  [arXiv:1308.5598 [hep-lat]].

\bibitem{Kikuchi:2014rla}
K.~Kikuchi and T.~Onogi, 
JHEP {\bf 1411}, 094 (2014) [arXiv:1408.2185[hep-th]].

\bibitem{Luscher:2011bx} 
  M.~L\"uscher and P.~Weisz,
  JHEP {\bf 1102}, 051 (2011)  [arXiv:1101.0963 [hep-th]].

\bibitem{Aoki:2014dxa} 
  S.~Aoki, K.~Kikuchi and T.~Onogi,
  JHEP {\bf 1504}, 156 (2015)
  [arXiv:1412.8249 [hep-th]].

  
\bibitem{Makino:2014sta} 
  H.~Makino and H.~Suzuki,
  PTEP {\bf 2015}, no. 3, 033B08
  [arXiv:1410.7538 [hep-lat]].

\bibitem{Makino:2014cxa} 
  H.~Makino, F.~Sugino and H.~Suzuki,
  arXiv:1412.8218 [hep-lat].

\bibitem{'tHooft:1974hx} 
  G.~'t Hooft,
  Nucl.\ Phys.\ B {\bf 75}, 461 (1974).
  
\end{thebibliography}
\end{document}